\documentclass[conference]{IEEEtran}
\IEEEoverridecommandlockouts

\usepackage{cite}
\usepackage{amsmath,amssymb,amsfonts}
\usepackage{graphicx}
\usepackage{textcomp}
\usepackage{xcolor}
\usepackage{comment}
\usepackage{url}

\usepackage{physics,algorithm}
\usepackage[noend]{algpseudocode}
\usepackage{amsthm}

\def\BibTeX{{\rm B\kern-.05em{\sc i\kern-.025em b}\kern-.08em
    T\kern-.1667em\lower.7ex\hbox{E}\kern-.125emX}}

\begin{document}



\title{Dynamic Routing in Space-Ground Integrated Quantum Networks\\
}


\author{
\IEEEauthorblockN{Tianjie Hu, Jindi Wu, and Qun Li}
\IEEEauthorblockA{\text{
Department of Computer Science, 
William \& Mary, 
Williamsburg, VA 23185, USA}
}}

\maketitle

\thispagestyle{plain}
\pagestyle{plain}

\newcommand{\RR}{\mathbb{R}}
\newcommand{\CC}{\mathbb{C}}
\newcommand{\dsp}{\displaystyle}




\begin{abstract}
\textbf{ } Quantum networks emerge as fundamental frameworks for addressing various large-scale problems. There are two primary architectures: space-based quantum networks, which deploy satellites with free space channels to interconnect users, and ground-based quantum networks, which utilize optical fibers to interconnect users. In this paper, we explore space-ground integrated quantum networks that incorporate both satellites and optical fibers into the infrastructure. This integrated network features three forms of communication: using only free space links, only ground links, or a hybrid usage of free space and ground links. We formulate the routing problem in space-ground integrated quantum networks as an integer programming and propose two solutions: using a linear relaxation and a greedy algorithm. The linear relaxation algorithm allows timely scheduling of additional entanglement purification, whereas the greedy algorithm enables quick scheduling. Simulation results demonstrate their effective balancing between network throughput and communication fidelity.
\end{abstract}


\section{Introduction}

Quantum networks provide a foundational framework for addressing large-scale problems such as distributed quantum machine learning, quantum key distribution, and clock synchronization. Within quantum networks, quantum communication is conducted via quantum teleportation, utilizing shared entanglement between end users. Recent technology advancements have enabled long-distance quantum communications by implementing multiple repeater nodes within networks to connect distant users. For each long-distance communication, pairwise entanglements are first established between adjacent users or repeaters, followed by quantum swapping to merge a chain of entanglements into a single entanglement between users. However, both entanglement generation and swapping are probabilistic processes, posing significant challenges for distant quantum communication. Therefore, efficiently and reliably distributing shared entanglements across the network is crucial, leading to the development of two promising architectures for quantum networks.

Space-based quantum networks~\cite{panigrahy2022optimal, chang2023entanglement, pirandola2021satellite}
are the most popular architecture for quantum networks due to their global-scale coverage and robust connections. In this architecture, user nodes are ground stations, and repeater nodes are typically low-Earth orbital satellites. Optical channels, established as free space links between ground stations and satellites, facilitate the distribution of entangled qubit pairs. However, since satellites are constantly moving, established optical channels may become invalid, and new ones may become available. Consequently, all connections within space-based quantum networks need frequent updates. This observation, combined with the probabilistic nature of entanglement generation, makes routing highly unpredictable and complex.

Another popular architecture for quantum networks is ground-based quantum networks~\cite{hu2023surfacenet, hu2024quantum}. In this architecture, both users and repeaters are ground stations, with repeater ground stations denoted as switches. Users and switches are interconnected via optical fiber cables. These physical optical fibers, serving as ground links, have fixed positions and connections once constructed, allowing for efficient scheduling and easy optimization of routing. However, the quality of communication through optical fibers decays exponentially with distance, necessitating a large number of repeaters within the network. Consequently, ground-based quantum networks are commonly limited to local settings.

\begin{figure}[t]
\centering
\includegraphics[scale=.54]{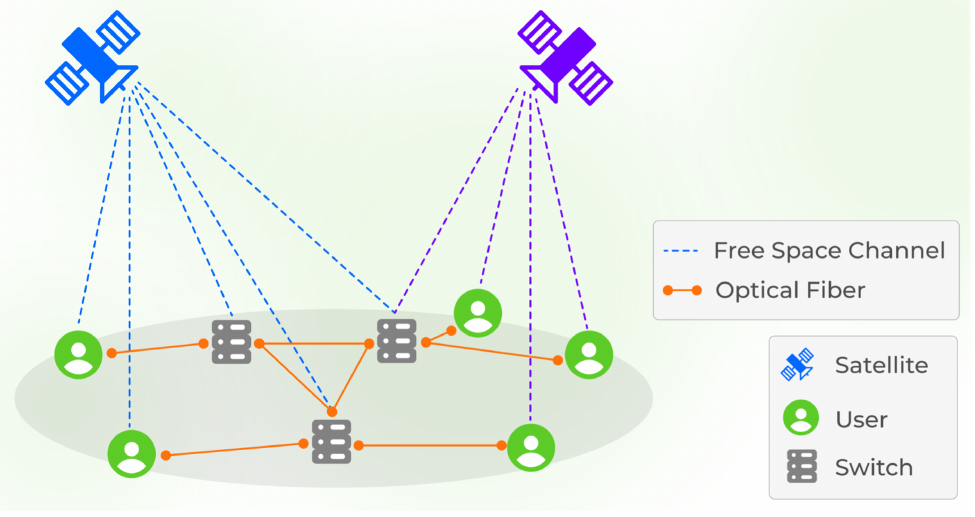}
\caption{\label{fig:network} \text{Space-ground integrated quantum network.} Users communicate with each other through satellite-based free space links and fiber-based ground links.}
\end{figure}

Note above that solely utilizing satellites with free space links may complicate entanglement distribution, while solely utilizing switches with ground links may restrict the operational scale of networks. Thus, in this paper, we investigate the architecture of a space-ground integrated quantum network, which incorporates both satellites and switches into a single network as depicted in Fig.~\ref{fig:network}. In our proposed space-ground integrated quantum network, three forms of communication, as shown in Fig.~\ref{fig:network_com}, can be conducted: using only free space links, only ground links, or a hybrid usage of free space and ground links. 
We evaluate the performance of our proposed network based on network throughput and average communication fidelity. Throughout this paper, \textit{throughput} is defined as the number of concurrent communications, and \textit{fidelity} is defined as the probability of communication occurring without any errors.
Our main contributions are summarized as follows:

\begin{itemize}
\item We present a comprehensive design for a space-ground integrated quantum network, effectively incorporating satellites and switches into its infrastructure. This network supports three forms of communication: using only free space links, only ground links, or a hybrid usage of free space links and ground links.
\item We formulate the routing protocol for our proposed space-ground integrated quantum network as an integer programming problem. And we present two solutions for solving the routing problem: using a linear relaxation algorithm and a greedy algorithm.
\end{itemize}

\section{Related Work}

Quantum networks emerge as fundamental frameworks for solving large-scale problems. Current physical implementations of quantum networks are still in their early stages, constrained by contemporary technology of qubit processing and storage, as well as their high costs. Efforts have been directed towards the theoretical design of quantum networks, including the mainstream space-based quantum networks~\cite{panigrahy2022optimal, chang2023entanglement, pirandola2021satellite} and ground-based quantum networks~\cite{hu2023surfacenet, hu2024quantum}. This paper explores the potential of space-ground integrated quantum networks. So far, plenty works~\cite{liu2018space} have been conducted in classical space-ground integrated networks, focusing on protocol design and network security. However, few works~\cite{bakker2024best} have been conducted in the quantum domain, and are primarily focused on communication rather than routing.

\section{Preliminaries}

\subsection{Free space and ground connections}

Communications in space-ground integrated quantum networks are conducted through free space links and ground links. Free space links connect satellites with ground stations, commonly implemented as optical channels that propagate photons through empty space. Meanwhile, ground links interconnect ground stations, typically implemented as optical fibers that propagate photons through constructed cables. 
In practice, communications between users may span long distances, necessitating the implementation of intermediate nodes, denoted as \textit{repeaters}, between senders and receivers. Repeaters are equipped with quantum memory and continuously generate entanglement pairs. These generated entanglements are then shared with its adjacent nodes via free space links or ground links. Repeaters can also perform quantum purification to convert multiple low-quality entanglements into a single high-quality entanglement, and quantum swapping to combine a chain of entanglements shared by adjacent nodes along the routing path into a single entanglement shared by end users.

\begin{figure}[t]
\centering
\includegraphics[scale=.39]{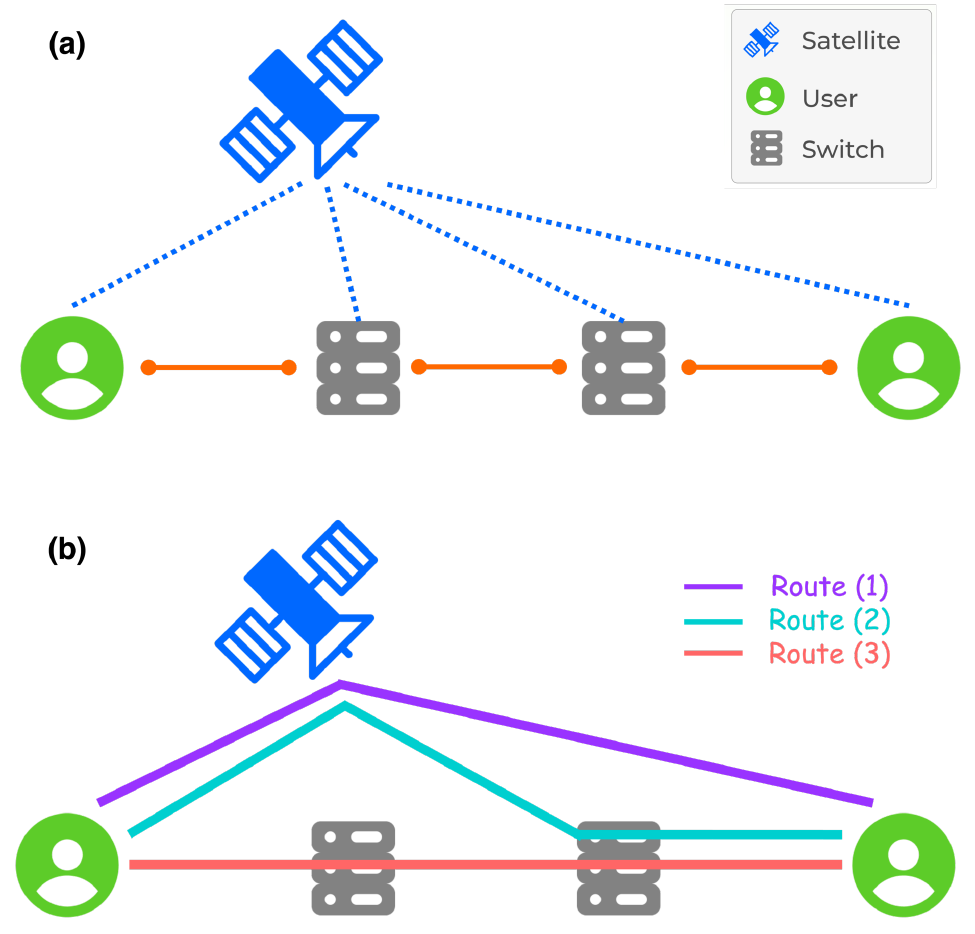}
\caption{\label{fig:network_com} (a) \text{Connections} between two users in a space-ground integrated quantum network. (b) \text{End-to-end communication} between the two users, where routing paths can be conducted through (1) free space links, (2) a hybrid usage of free space and ground links, or (3) ground links. Actual choices are made by the routing protocol in the network.}
\end{figure}


In space-ground integrated quantum networks, users are commonly located at ground stations. \textit{Satellites} serve as free space repeaters that distribute entanglements via free space links, and several ground stations, denoted as \textit{switches}, serve as ground repeaters. Example network connections are illustrated in Fig.~\ref{fig:network_com}(a). And as shown in Fig.~\ref{fig:network_com}(b), end-to-end communications in space-ground integrated quantum networks occur in three forms. Notably, communications that rely solely on free space links, as seen in Route (1), generally involve only a single satellite, as free space optical channels can be established over global distances. In contrast, communications that rely solely on ground links, as seen in Route (3), generally traverse multiple switches since ground optical fiber channels typically span only hundreds of kilometers. This paper also explores a hybrid communication, illustrated as Route (2), where both free space links and ground links are used within a single communication. 
This hybrid communication can effectively exploit the differing reliability and entanglement generation rates in optical channels and optical fibers.

\subsection{Satellite constellations}

Unlike ground stations that are stationary at geographic locations, satellites are constantly moving and need to be arranged in \textit{constellations} to avoid collisions. A popular choice is the Walker Delta constellation, which is a symmetric arrangement consisting of $n$ satellites evenly distributed across $p$ orbital planes. The ascending nodes (intersections between orbital planes and the reference plane, which is typically the equator) of these planes are uniformly spaced along the equator. Each Walker Delta constellation configuration is commonly denoted as $\delta:n/p/f$, where $n$ represents the number of satellites, $p$ indicates the number of orbital planes, $\delta$ specifies the inclination angle between each orbital plane and the reference plane, and $f$ describes the phasing between satellites in adjacent planes.


As each satellite is constantly moving, it may no longer be able to maintain optical channels with all ground stations, and the quality of the established optical channels varies depending on their relative locations and distances.
To establish a valid optical channel, the elevation angle $\theta^s_g$ between satellite $s$ and the horizon at ground station $g$ needs to exceed the elevation angle threshold $\theta^{th}$. And there are two choices for distributing photons through each optical channel, known as \textit{uplink} and \textit{downlink}. Uplinks occur when photons are sent from ground stations to satellites, while downlinks occur in the opposite direction. Generally, downlinks exhibit better performance in terms of loss rates under low signal-to-noise ratios (SNR)~\cite{pirandola2021satellite}. Thus, a double downlink architecture has become a popular implementation choice for each satellite, where all entanglements through free space links are established from satellites to ground stations.


\subsection{Noise and Loss model} \label{sec:bg}

Entanglement distribution through optical channels or optical fibers is susceptible to various noise sources. Commonly raised errors include depolarization errors and erasure errors. Depolarization errors occur when the quantum state of an entanglement qubit is unintentionally altered by noise, such as unfiltered background photons or environmental factors. Erasure errors, on the other hand, occur when an entanglement qubit is lost during transmission. Other quantum errors, such as decoherence errors, cross-talk errors, idling errors, and more, can also arise. Like depolarization and erasure errors, these can result in either unintended changes to the entanglement or its complete loss.
Correspondingly, we evaluate the quality of each optical channel or optical fiber in terms of its \textit{fidelity}, defined as its probability of preserving the integrity of each entanglement qubit transmitted through it, preventing the qubit from any errors. Fidelity is commonly measured within $[0,1]$, with higher fidelity indicating better quality.

The fidelity of each optical fiber can be directly determined once the cable is constructed. However, since satellites are in constant motion, the fidelity of each optical channel needs to be frequently calculated and updated in the routing protocol in a timely manner.
Let $f_{s,g}$ denote the free space distance between satellite $s$ and ground station $g$, and $h_{s,g}$ represent the distance between the atmospheric boundary and ground station $g$ along $f_{s,g}$. Additionally, let $d^T_{s}$ and $d^R_{g}$ be the diameters of the transmitter and receiver telescopes, respectively at satellite $s$ and ground station $g$, both operating at wavelength $\lambda$. The \textit{channel transmissivity} $\eta_{s,g}$ of this optical channel can be computed using the below formula as presented in~\cite{panigrahy2022optimal}:
\begin{equation}
\eta_{s,g}= \frac{(\pi(d^T_{s}/2)^2)(\pi(d^R_{g}/2)^2)}{(\lambda*f_{s,g})^2}*e^{-\alpha h_{s,g}}
\end{equation}
where $\alpha$ is the atmospheric extinction coefficient. Based on the channel transmissivity, the capacity of this optical channel is calculated in~\cite{pirandola2021satellite} as $-\log_2(1-\eta_{s,g})$, while the relationship between entanglement fidelity and channel transmissivity is analyzed in~\cite{kumar2023success}. For ease of discussion in this paper, we assume that both the capacity and fidelity of each optical channel can be directly obtained by the routing protocol, similar to those of each optical fiber. 
\vspace{.06in}

The fidelity of each communication is then computed by multiplying the fidelity of all optical channels and fibers it traverses through. Note that as discussed later in Section~\ref{sec:form}, each satellite or switch may also introduce additional noise to communications. Meanwhile, various techniques exist to enhance communication fidelity. For instance, entanglement purification can be employed in satellites and switches by consolidating multiple low-quality entanglements into a single high-quality entanglement. For example, by consuming two entanglement pairs with fidelity $\rho_1$ and $\rho_2$, the result entanglement is with a new fidelity $\rho$ calculated in~\cite{pan2001entanglement} as:
\begin{equation}
\rho= \frac{\rho_1\rho_2}{\rho_1\rho_2+(1-\rho_1)(1-\rho_2)}
\end{equation}
However, entanglement purification comes with the drawback of increased traffic in networks, as the original two entanglements could support two communications instead of the current one. Therefore, it becomes a crucial design feature in the routing protocol to manage the frequency of entanglement purification operations within the network.
\section{Problem Formulation} \label{sec:form}

Given limited capacity at satellites and switches, and constrained entanglement generation across optical channels and fibers, we propose a centralized routing protocol aiming to maximize network throughput while ensuring high fidelity for each communication. Again, \text{throughput} is defined as the number of concurrent communications, and \text{fidelity} is defined as the probability of occurring without any errors. We make the following assumptions:
\begin{itemize}
\item The network is connected, ensuring the existence of at least one path between any two users.
\item Satellites, switches, optical channels, and optical fibers have finite capacities.
\item Fidelity of optical channels and fibers can be measured and remain constant during each routing round. Fidelity of optical channels may vary between rounds.
\end{itemize}
We consider two common errors in quantum networks: single-qubit depolarization errors and erasure errors. However, our method can be generally extended to account for any other types of quantum errors, by updating the fidelity of each optical channel and fiber accordingly.

\subsection{Routing Procedure}


Before each round of routing, the routing protocol collects requests from user nodes, and operating information from satellites and switches including their current capacities and entanglement status. It also updates the connectivity of each free space link. The fidelity of connection between an arbitrary satellite $s$ with a ground station $g$ at time $t$ is expressed as:
\begin{equation}
f_{s,g}(t)= \left\{
\begin{array}{ll}
      \text{correlated to $\eta_{s,g}(t)$}, & \theta^s_g(t) \ge \theta^{th}\\
      0.001, & \textit{otherwise} \\
\end{array} 
\right.
\end{equation}
where $\theta^s_g $ and $\eta_{s,g}$ are their elevation angle and transmissivity in between, and $\theta^{th}$ is the elevation angle threshold. Note the $0.001$ here represents a small non-zero number, for the convenience of mathematical formulation later.
The resulting graph, containing updated satellite positions and optical channels, along with unchanged ground stations and optical fibers, is denoted as the \textit{routing graph}. 
Both satellites and switches are treated as repeaters in this graph, serving as intermediate nodes along communication paths, yet they greatly differ in entanglement generation rates. Similarly, optical channels and fibers are both treated as edges, differing in fidelity and transmission rates. Thus, entanglement purification is typically more frequently scheduled for communications across optical fibers, particularly for long-distance connections.

\begin{table}[t]
\begin{center}
\caption{\label{table:notation}TABLE OF NOTATIONS USED IN ROUTING FORMULATION}
\begin{tabular}{|| l l ||} 
\hline\hline
 Notation & Definition \\ 
\hline\hline
 $\RR$ & The set of all repeaters \\
 $\mathbb{E}$ & The set of all edges (including ground-satellite edges)\\
 $\mathbb{ES}$ & The set of ground-satellite edges \\
\hline
 $C_e$ & Number of prepared entanglements across each edge $e\in E$ \\
 $f_{e}(t) $ & Fidelity at each edge $e\in \mathbb{ES}$ at time $t$\\
 $\mu_e $ & Noise at each edge $e\in\mathbb{E}$\\
 $p_e $ & Effect of entanglement purification at each edge $e\in\mathbb{E}$\\
 $C_r$ & Storage capacity in each repeater $r\in R$ \\
 $\sigma_r$ & Noise amendment at each repeater $r$\\
\hline
 $N^{th}$ & Noise threshold for each communication\\
 $\mathbb{K}$ & The set of all communication requests $k=[(s_k,d_k),m_k]$\\
 $m_k$ & Number of qubits in request $k$\\
\hline
 $Y_k$ & Integer variable of determining number of messages to\\
 & be transmitted for request $k$\\
 $x^k_e$ & Integer variable of determining number of entanglements\\
 & consumed at edge $e$ for request $k$ \\
 $\phi^k_e$ & Integer variable of determining number of entanglements\\
 & for extra purification consumed at edge $e$ for request $k$ \\
 $\alpha^k_r$ & Integer variable of determining number of swapping in\\
 & repeater $r$ for request $k$ \\
 \hline
\end{tabular}
\end{center}
\end{table}

\subsection{Mathematical Formulation}

To provide a formal mathematical formulation of our routing protocol, we define key terms as outlined in Table~\ref{table:notation}. Each communication request $k$ is denoted as $[(s_k,d_k),m_k]$, where $s_k$ and $d_k$ represents the sender and receiver nodes, and $m_k$ indicates number of qubits in the message. Decision variables in our formulation consist of four sets of integer variables: $Y_k$ to decide whether each request will be scheduled or partially scheduled, $x_e^k$ to decide routes for each request, and $\phi^k_e$ to decide whether extra purification is needed for each route. We also include $\alpha_r^k$ to record the corresponding repeater nodes traversed by each potential routing path.

Estimating fidelity for each potential routing path involves a sequence of multiplications, resulting in a non-linear problem formulation. Thus, we transform the fidelity at each edge into its corresponding \textit{noise}, to make the routing problem more tractable. The noise at each edge is calculated as $\mu_e = \log(1/\gamma_e)$, where $\gamma_e$ is its fidelity measured in $[0,1]$. This transformation allows noise to accumulate through addition rather than multiplication, with lower values indicating better quality. Correspondingly, we model the effect of extra purification scheduled at each edge as a reduction $p_e$ on the accumulated noise. We denote $\kappa_e$ as the number of purification required to achieve a 0.99 fidelity at each edge, and calculate the effect of each purification $p_e=\frac{\mu_e}{\kappa_e}$ so that it scales linearly.
In addition, at each repeater, entanglement swapping has a non-trivial failure rate and may introduce additional noise. Thus, we assign a noise amendment $\sigma_r$ to each repeater node, ensuring that the routing protocol discourages excessive use of intermediate nodes that can degrade the overall fidelity of the communication.
\vspace{.06in}

Our integer programming for routing states as follows. The objective function is to sum the number of concurrent communications, with the goal of maximizing network throughput:

\begin{equation}
\max \sum_{k\in \mathbb{K}} Y_k
\end{equation}
with the following constraints:
\begin{equation*}
\label{eq:c1}
\begin{aligned}
&\ Y_k\in [0,m_k] \textbf{ , } x^k_e\ge 0 \textbf{ , } \phi^k_e\ge 0 \textbf{ , } \alpha^k_r\ge 0 &\ &\\[1ex]
&\ \sum_{(i,s_k)\in E} x^k_{(i,s_k)}+ \sum_{(d_k,j)\in E} x^k_{(d_k,j)}= 0 & \forall_{k\in \mathbb{K}} &\\ 
&\ \sum_{(s_k,j)\in E} x^k_{(s_k,j)}= \sum_{(i,d_k)\in E} x^k_{(i,d_k)} = Y_k & \forall_{k\in \mathbb{K}} &\\
&\ \sum_{(i,r)\in E} x^k_{(i,r)} = \sum_{(r,j)\in E} x^k_{(r,j)}=\alpha^k_r & \forall_{ r\in\mathbb{R}\textbf{, } k\in \mathbb{K}} &\\
&\ \mu_e= \log(1/f_{e}(t)) \text{ , } p_e= 0 & \forall_{e={(s,g)}\in \mathbb{ES}} &\\[.5ex]
&\ \sum_{k\in\mathbb{K}} x^k_e+\phi^k_e \le C_e & \forall_{e\in\mathbb{E}} &\\
&\ \mu_e*x^k_e-p_e*\phi^k_e\ge 0 & \forall_{ e\in E\textbf{, } k\in \mathbb{K}} &\\
&\ \sum_{k\in\mathbb{K}} \sum_{(i,r)\in E} x^k_{(i,r)} \le C_r & \forall_{r\in\mathbb{R}} &\\
&\ \sum_{e\in E}(\mu_e x^k_e-p_e \phi^k_e)
+ \sum_{r\in \RR} \sigma_r  \alpha^k_r \le N^{th}*Y_k 
& \forall_{k\in \mathbb{K}} &\\
\end{aligned}
\end{equation*}
The first four lines of constraints formally set up a flow network, including the \textit{initialization} and \textit{termination} constraints that apply to sender and receiver nodes, and the \textit{conservation} constraints that apply to repeater nodes. The fifth line updates the potential noise within all free space links, as each satellite is constantly in motion and the connectivity of these links may vary over time. The effect of purification at each free space link is set to 0, since optical channels are commonly established with high fidelity, making extra purification unnecessary.

The next three lines are the \textit{capacity} constraints that ensure the computing resources within the network are not over-consumed. Line 5 sums the number of entanglements scheduled to be consumed across each edge, ensuring they do not exceed the number of prepared ones. Line 6 prevents excessive entanglement purification scheduled for each communication to conserve computing resources. Line 7 sums the number of entanglements scheduled for storage in each repeater, ensuring they do not exceed the memory capacity at each location. Note that each $\phi^k_e$ does not need to be stored in quantum memory as it has already been consumed for purification.

The last line contains the \textit{noise} constraints that ensure all communications maintain high fidelity, calculated as summing the noise from all edges and repeaters along the routing path. If extra purification is scheduled for an edge, the noise induced in that edge will be correspondingly reduced. The summation of noise for each path must be below threshold $N^{th}$ to ensure high fidelity for each communication. As demonstrated later in Section~\ref{sec:eval}, fine-tuning $N^{th}$ allows for balancing between communication fidelity and network throughput.

\section{Proposed Solutions}

Note that in the above formulation, $C_e,C_r,\mu_e,p_e,\sigma_r,m_k$ are collected from the network, and $N^{th}$ is a configurable parameter, ensuring that the above constraints remain linear. However, since all decision variables $Y_k,x^k_e,\phi^k_e,\alpha^k_r$ are integers, the above integer programming does not guarantee a polynomial-time solution. In this section, we propose two solutions for solving the above routing problem.

\subsection{Linear Relaxation Algorithm}

As a common practice for solving integer programming problems, \textit{linear programming relaxation} can be applied by removing the integer constraints for all integer variables, allowing them to be any real numbers. By solving the relaxed version of the problem, the optimal solutions obtained are likely to be non-integers. Then a rounding process is necessary to convert these solutions back into integer values. Thus, linear programming relaxation does not guarantee a truly optimal solution to the original integer programming problem but provides a near-optimal approximation.
For instance, to solve the above routing problem, we treat all decision variables $Y_k, x^k_e, \phi^k_e, \alpha^k_r$ in our formulation as real numbers. Note that the rounding process requires additional efforts, as it may violate both the conservation and capacity constraints in our formulation. Thus, when rounding the optimal solution, we need to decide whether to round each decision variable up or down accordingly to ensure all constraints are still met.

\begin{algorithm}[t]
\caption{Greedy Algorithm}\label{alg:gre}
\textbf{Input:} routing graph $G=\{V,E,W\}$, set of requests $\{k_i\}$,
\\ \hspace*{\algorithmicindent}\hspace*{\algorithmicindent}
capacity of each node $C_v$, capacity of each edge $C_e$, 
\\ \hspace*{\algorithmicindent}\hspace*{\algorithmicindent}
noise at each node $\mu_v$, noise at each edge $\mu_e$, 
\\ \hspace*{\algorithmicindent}\hspace*{\algorithmicindent}
noise threshold $N^{th}$
\\
\textbf{Output:} scheduled routing paths $\{R_{k_i}\}$ 
\begin{algorithmic}[1]
\State Construct subgraph $G_1$ of $G$ with only ground links
\State Construct subgraph $G_2= G$
\State Construct subgraph $G_3$ of $G$ with only free space links
\State Initialize sets $P_{G_1}, P_{G_2}, P_{G_3}$ as empty sets
\For{each request ${k_i}=\{s_{i},d_{i},m_{k_i}\}$}
    \For{${G_x}$ in $\{{G_1}, {G_2}, {G_3}\}$}
        \State Find shortest path $p_{x,{k_i}}$ between $s_{i},d_{i}$ in $G_x$ 
        \If{noise accumulated in $p_{x,{k_i}}$ is below $N^{th}$}
            \State Count required entanglements in $p_{x,{k_i}}$ as $c_{x,{k_i}}$
            \State Add tuple $(p_{x,{k_i}},c_{x,{k_i}},m_{k_i})$ into $P_{G_x}$
        \EndIf
    \EndFor
\EndFor
\For{$P_{G_x}$ in $\{P_{G_1}, P_{G_2}, P_{G_3}\}$}
    \State Sort $P_{G_x}$ in increasing order of $c_{x,{k_i}}$
    \State \hspace*{\algorithmicindent}WLOG, assume $c_{x,{k_1}}\le c_{x,{k_2}}\le \dots \le c_{x,{k_n}}$
        \For{${k_i}:= {k_1} \textbf{ to } {k_n}$}
        \State Find largest integer $\alpha\le m_{k_i}$ that capacity required \hspace*{\algorithmicindent}\hspace*{\algorithmicindent}by $\alpha$ turns of $p_{x,{k_i}}$ can be satisfied, and remove \hspace*{\algorithmicindent}\hspace*{\algorithmicindent}the required capacity from $C_v$ and $C_e$ 
        \State Update $m_{k_i}=m_{k_i}-\alpha$ in $P_{G_1}$, $P_{G_2}$ and $P_{G_3}$
        \State Add $(p_{x,{k_i}},\alpha)$ into $R_{k_i}$
    \EndFor
\EndFor
\end{algorithmic}
\end{algorithm}

\subsection{Greedy Algorithm}

Another approach is to design a \text{greedy algorithm} for solving the routing problem. Our proposed greedy approach is outlined in Algorithm~\ref{alg:gre}. Input to the algorithm includes the routing graph $G$, which is a weighted graph with users and repeaters as nodes, and free space links and ground links as edges. The weight of each edge is its noise as calculated in our integer formulation. Capacity and noise are directly collected from the network, with the noise at each node corresponding to the noise amendment in our integer formulation. The noise threshold $N^{th}$ is configurable, similar to the noise threshold in our integer formulation, and can be fine-tuned to balance between communication fidelity and network throughput.
\vspace{.06in}

The first step of our proposed greedy algorithm involves constructing two subgraphs, $G_1$ and $G_3$, respectively comprising only ground links or free space links. Next, for each request, three potential routing paths are constructed using ground links, a hybrid usage of free space and ground links, or free space links. The noise of each constructed path is computed and is required to be below the noise threshold. 
Next, the algorithm sequentially goes through each set ($P_1$, $P_2$, and then $P_3$) of routing paths. Within each set, the algorithm greedily selects the paths that consume the fewest entanglement pairs along the path and include them into the scheduling.

Notably, outputs from Algorithm~\ref{alg:gre} are not guaranteed to be optimal, but this algorithm can be very quick in scheduling the routing. Denote $V$ as the number of nodes, $E$ as the number of edges, and $K$ as the number of requests. Constructing subgraphs requires $\mathcal{O}(V+E)$, computing minimal weighted paths can be solved using \textit{Dijkstra's} in $\mathcal{O}(K*V^2)$, sorting requires $\mathcal{O}(K\log{K})$, and updating the capacities requires $\mathcal{O}(K*(V+E))$. Thus, as $\mathcal{O}(K)\subseteq \mathcal{O}(V^2)$, the running time of Algorithm~\ref{alg:gre} is bounded by the shortest paths calculation and exhibits a worst-case time complexity of $\mathcal{O}(K*V^2)$.

\section{Evaluation} \label{sec:eval}

We conduct simulations to evaluate the performance of our space-ground integrated quantum network with our proposed routing protocols. To the best of our knowledge, there has not been any routing protocol proposed for scheduling simultaneous communication requests in space-ground integrated quantum networks. Thus, comparison is conducted between the two routing algorithms proposed in this paper. In the following, we denote our linear relaxation algorithm as \textbf{Linear} and our greedy algorithm as \textbf{Greedy}. We evaluate the performance using three metrics: \textit{fidelity}, \textit{latency}, and \textit{throughput}. Fidelity measures the success rate of each communication, while latency quantifies the waiting time for each communication. Both fidelity and latency are computed as averages over all communications executed in the network. Throughput is calculated as the ratio of the number of executed communications to the total number of requested communications.


Both routing algorithms are tested across three network scenarios: networks with abundant allocations of satellites and switches, networks with sufficient allocations, and networks with insufficient allocations. All experiment networks are randomly generated. The topology of ground stations is generated using the \textit{Barabasi-Albert} model, with the most connected nodes chosen to be switches. Satellites are randomly assigned to different locations within the generated network and connected to local ground stations. Fidelity of each free space link is randomly assigned within range $[0.9, 1]$, and fidelity of each ground link is randomly assigned within range $[0.75, 1]$. The success rate for each entanglement swapping is fixed at $95\%$. During each experiment trial, capacities for nodes and edges are randomly assigned, and communication requests are generated between random pairs of users.

\begin{figure}[t]
\includegraphics[scale=.58]{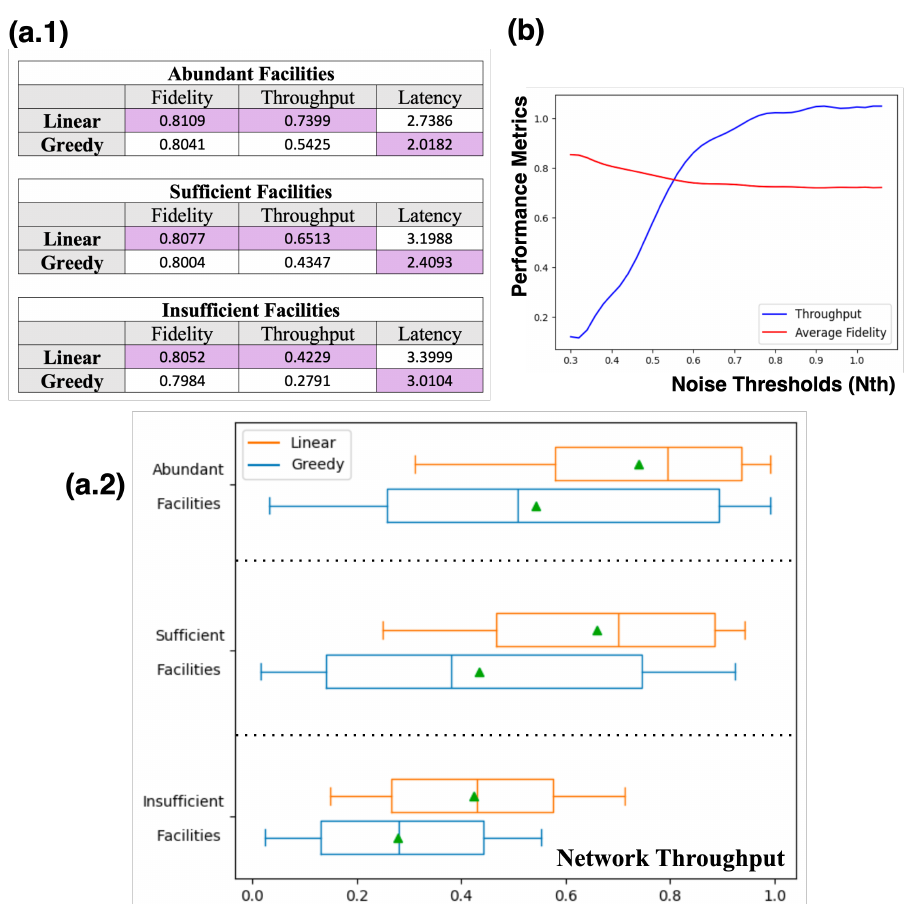}
\caption{\text{(a.1) Comparison between Linear and Greedy} in different network scenarios over three metrics: fidelity, throughput, and latency. \text{(a.2)} Detailed comparison on network throughput. \text{(b) Performance of Linear} with respect to its configurable parameter Noise threshold.}
\label{fig:eval_r}
\end{figure}

From table (a.1) in Fig.~\ref{fig:eval_r}, the two routing protocols exhibit similar average fidelity across all three network scenarios, demonstrating effective control of communication fidelity by the noise threshold in Linear and the fidelity threshold in Greedy. However, Linear outperforms Greedy in terms of network throughput, enabling more communications to be conducted simultaneously. Meanwhile, Linear demonstrates worse average latency compared to Greedy, primarily due to the latter's greedy nature, which prioritizes scheduling communications with shorter distances. A detailed comparison of network throughput is presented in (a.2). Notably, the performance of Greedy appears less stable than that of Linear, with significant variation among different trials and more instances of low network throughput.
And in Fig.~\ref{fig:eval_r} (b), we investigate the impact of manipulating the noise threshold in Linear. As previously mentioned, the noise threshold can be adjusted to strike a balance between communication fidelity and network throughput. As in the plot, a higher noise threshold allows for more communications with lower fidelity, resulting in higher network throughput but lower average fidelity. Conversely, a lower noise threshold prioritizes communication quality, leading to reduced network throughput but higher average fidelity. Similar trade-offs between fidelity and throughput are also observed in experiments conducted with Greedy.

\section{Conclusion}

In this paper, we presented a space-ground integrated quantum network along with our proposed routing protocol. Within the network, satellites and switches are deployed as repeater nodes, with optical channels serving as free space links and optical fiber cables serving as ground links. Three forms of communication are facilitated: using only free space links, only ground links, or a hybrid usage of both. We formulated the routing problem as an integer programming and proposed two solutions: using linear relaxation or a greedy algorithm. The linear relaxation approach allows for the prompt scheduling of additional entanglement purification but necessitates adjustments for rounding. On the other hand, the greedy approach offers straightforward solutions with efficient implementation, but it exhibits unstable performance. Simulations demonstrated that our routing protocols can effectively balance between network throughput and average communication fidelity.

\section*{Acknowledgments}
The authors would like to thank all the reviewers for their helpful comments. 
This work was supported by the Commonwealth Cyber Initiative (\url{cyberinitiative.org}).
\vspace{.08in}

\bibliographystyle{ieeetr}
\bibliography{sfnet}

\end{document}